\documentstyle[sprocl,epsf]{article}

\def\bA{{\bf A}}
\def\bB{{\bf B}}
\def\bE{{\bf E}}
\def\bL{{\bf L}}
\def\bS{{\bf S}}
\def\br{{\bf r}}
\def\bhr{{\bf\hat r}}
\def\bp{{\bf p}}
\def\bv{{\bf v}} 
\def\bhvp{{\bf\hat v}_\perp}
\def\openone{\leavevmode\hbox{\xipt1\kern-3.8pt\xiipt1}}

\begin{document}
\vglue-.5in

\font\fortssbx=cmssbx10 scaled \magstep1
\hbox to \hsize{
%\special{psfile=uwlogo.ps hscale=6000 vscale=6000 hoffset=-12 voffset=-2}
%\hskip.25in \raise.1in
\hbox{\fortssbx University of Wisconsin - Madison}
\hfill$\vcenter{\hbox{\bf MADPH-96-964}
                \hbox{September 1996}}$ }

\bigskip

\title{CONFINEMENT, SPIN, AND QCD\,\footnote{Adapted from talks given at the {\it Quarkonium Physics Workshop}, University of Illinois at Chicago, June 13--15, 1996, and at the {\it International Conference on Quark Confinement and the Hadron Spectrum II}, Como, Italy, June 26--29, 1996.}}

\author{M. G. OLSSON}

\address{Physics Department, University of Wisconsin, 1150 University Ave.\\
Madison, WI 53706, USA}

\maketitle

\abstracts{The experimental and theoretical evidence
relating spin dependence and long range confinement is
reviewed. One of the simplest confinement ideas, the dynamic
electric flux tube picture of Buchm\"uller, can be exploited to
yield a complete effective Hamiltonian. In addition to pure 
Thomas spin-orbit splitting the other relativistic corrections
are calculated.  It is shown
that our effective Hamiltonian is identical to the mechanical
relativistic flux tube model and gives relativistic corrections in agreement with QCD predictions.}

\section{Introduction}

The linear growth in energy with increasing quark and anti-quark 
separation has long been accepted as a natural prediction of QCD.\cite{wilson}
On the other hand, the quark's orientational energy 
has been a matter of continuing discussion.  It is clear that the long
range spin interaction cannot be similar to the short range gluon
exchange since this would imply a large hyperfine splitting in the $p$-wave
charmonium states which is at least an order of magnitude larger than
observed.\cite{como94} About twenty years ago Schnitzer\cite{schnitzer} considered the ratio of $p$-wave $\chi$-state masses,
\begin{equation}
    R= {\chi_2-\chi_1\over\chi_1-\chi_0} \simeq 0.5\rm\ (exp.)
\end{equation}
If the interaction potential were of the Lorentz vector type $R>0.8$ but it was observed that
a long range Lorentz scalar interaction would 
decrease the ratio $R$ and successful phenomenological models could be constructed. The improvement of course is due to the negative spin-orbit energy from scalar confinement whereas
the vector case is positive.  This observation ignited a long tradition
of assuming scalar confinement. The $R$ argument assumes the spin dependence
is given by the Breit-Fermi Hamiltonian in the ladder approximation. The
argument also assumes that virtual coupling to open flavor channels
can be ignored. It should also be pointed out that the justification of the hypothesis of scalar confinement  from a fundamental QCD point of view
has never been very satisfactory.

The discussion of long range spin dependence was placed on a firm foundation
by the low velocity expansion of the Wilson loop minimal area law by
Eichten and Feinberg.\cite{eichten}  For one static and one slowly moving quark
the spin-orbit energy is given in this formalism by
\begin{equation}
H_{\rm s.o.} \simeq {1\over 2m^2r} \left({dV\over dr} + 2{dV_1\over dr}\right) \bf S\cdot L \,,
\label{eq:eich-fein}
\end{equation}
where $V$ is the static potential and $V_1$ is the long range potential. It was then observed by Gromes\cite{gromes} that there is a further constraint on these potentials,
\begin{equation}
{dV\over dr} = {dV_2\over dr} - {dV_1\over dr} \,,
\label{eq:gromes}
\end{equation}
where $V_2$ is the short range interaction.
At large quark separations $V_2\ll V_1$ and with linear confinement $V\to ar$
the spin-orbit energy (\ref{eq:eich-fein},\ref{eq:gromes}) becomes
\begin{equation}
H_{\rm s.o.} \simeq -{a\over 2m^2 r}  \bf S\cdot L \,,
\label{eq:gromes-so}
\end{equation}
which is exactly the Thomas spin-orbit energy term.  The ``Gromes relation" (\ref{eq:gromes}) thus provides
further evidence for a pure Thomas spin-orbit interaction at large separations.
Some doubt has been cast on the original derivation in recent work by Williams.\cite{ken} We also await a careful study of the $B_c$ $p$-wave potential by lattice simulation.

A simple and attractive picture relating confinement and spin-dependence was proposed by Buchm\"uller.\cite{buchmuller} In this picture the quarks carry the color field around as they move and for large separations the the field collapses to a  purely electric flux tube. There are two immediate conclusions one can draw in this limit. First, since the color magnetic moments of the quarks are decoupled there can be no long-range hyperfine or tensor interactions. Second, since there is no magnetic field the only spin-orbit interaction is kinematic, known as the ``Thomas" spin-orbit energy. In this talk I will extend the Buchm\"uller picture considerably to show that this simple concept leads to a compete theory and that it is consistent with other results based in more formal applications of QCD.

\section{A Review of Buchm\"uller's Argument}

Although QCD is the proper theory of quark interaction, it is not unreasonable to discuss some questions within the context of electrodynamics. The flux-tube field configuration certainly depends crucially on the nonabelian nature of QCD while the interactions of the quarks with the fields does not in our case. This is because we assume a flux-tube field in which only one component of the field strength $F_{\mu\nu}$ (i.e., the radial color electric field) is non-zero in a co-rotating frame. We may thus gauge transform $F_{\mu\nu}$ to an abelian subgroup of SU(3).

We consider a charged particle of spin \bS\ and gyromagnetic ratio $g$. Its classical motion in electric and magnetic fields {\bf E} and {\bf B} is given by the Thomas equation.\cite{thomas} If the particle is viewed from a frame in which its motion is only radial (the co-rotating frame) the fields in this frame ($\bf E'$ and $\bf B'$) are given by
\begin{equation}
\begin{array}{rcl}
\bf E &=& \gamma_\perp (\bE' - \bv_\perp \times \bB') \,,\\
\bf B &=& \gamma_\perp (\bB'_\perp + \bv_\perp \times \bE')\,.
\end{array}
\label{eq:lorentz}
\end{equation}
The spin-orbit energy\cite{jackson} in terms of co-rotating fields is
\begin{equation}
H_{\rm s.o.} = -{1\over m}{\bf S} \cdot \left({g\over 2\gamma_\perp} {\bf B'} + {1\over\gamma_\perp+ 1} \bf v_\perp \times E' \right) \,,
\label{eq:so-energy}
\end{equation}
where $\gamma_\perp^{-2} = 1 - v_\perp^2$ and the charge has been absorbed into the fields. For an electric flux tube $\bf B' \equiv 0$ at every point along the flux tube (and also in the quark rest frame). The co-rotating electric field is given by
\begin{equation}
{\bf E'} = -a\bf\hat r
\end{equation}
in the flux tube limit. For a slowly moving quark ($v\ll 1$) the spin-orbit energy (\ref{eq:so-energy}) becomes
\begin{equation}
H_{\rm s.o.} \simeq -{a\over 2m^2r} \bf S\cdot L \,,
\end{equation}
where ${\bf L} \simeq m {\bf r\times v}$. We note this is the same as the Gromes result (\ref{eq:gromes-so}). In what follows, the validity of the Buchm\"uller picture is assumed as is the consequent Thomas-type spin-orbit interaction.

\section{Equivalent Laboratory Fields}

For simplicity we will assume here that the anti-quark is infinitely heavy and fixed at the origin. A straightforward extension to a general meson would consider two collinear flux-tube segments in the CM system with CM at the coordinate origin. As in Fig.~1 an arbitrary point on the tube is $\bf r$, the quark's position is ${\bf r}_Q$, and the momentum conjugate to ${\bf r}_Q$ is ${\bf p}_Q \equiv -i\hbox{\boldmath$\nabla$}_Q$.

\begin{figure}[t]
\centering
\epsfxsize=2.75in\hspace{0in}\epsffile{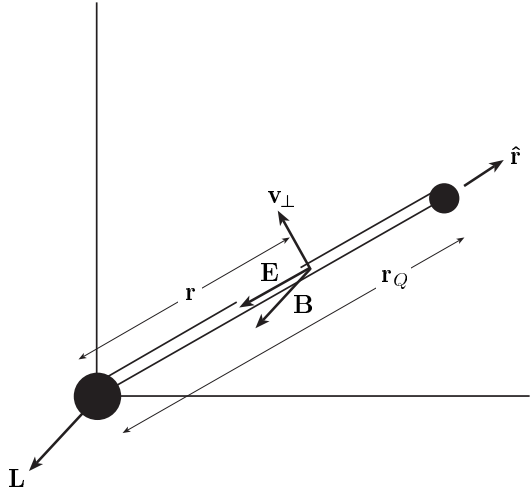}
\medskip
\caption{Flux tube with static $\bar Q$ at the coordinate origin. The moving quark is located at~$\br_Q$. In a non-rotating frame the fields are \bE\ and \bB\ at the point \br\ on the flux tube.}
\end{figure}

As we have discussed, the fields in the co-rotating frame in Buchm\"uller's picture are
\begin{equation}
\begin{array}{rcl}
{\bf E'} &=& -a\bf\hat r \,,\\
{\bf B'} &=& 0\,.
\end{array}
\label{eq:buch-fields}
\end{equation}
We now use the Lorentz transformations (\ref{eq:lorentz}) to find the lab fields {\bf E} and {\bf B} which (at the same location on the flux tube) are equivalent to the co-rotating fields (\ref{eq:buch-fields}). They are
\begin{equation}
\begin{array}{rcl}
{\bf E} &=& \gamma_\perp {\bf E'} = -a\gamma_\perp \bf\hat r\,,\\
{\bf B} &=& a\gamma_\perp v_\perp (\bf\hat r \times \hat v_\perp)\,.
\end{array}
\end{equation}
For a straight flux tube the magnitude of the perpendicular velocity is
\begin{equation}
v_\perp = {r\over r_Q} v_{Q\perp} \,,
\label{eq:vperp}
\end{equation}
where $v_{Q\perp}$ is the component of quark velocity perpendicular to the flux tube.

We next proceed to define a four vector potential at points along the tube in the lab system. The field vector potential is defined by
\begin{equation}
\begin{array}{rcl}
\bB(\br, \br_Q, \bp_Q) &\equiv& \hbox{\boldmath$\nabla$}_{\br} \times \bA (\br,\br_Q,\bp_Q)\\
&=&\displaystyle a {r\over r_Q} v_{Q\perp}\> \gamma_\perp \!\left({r\over r_Q} v_{Q\perp}\right) \left(\bhr\times\bhvp\right) \,,
\end{array}
\label{eq:fvp}
\end{equation}
where we explicitly denote the spatial dependence of $\gamma_\perp(v_\perp)$ using (\ref{eq:vperp}).
We choose the gauge $\bA = A_\perp(r) \bhvp$, which implies $\hbox{\boldmath$\nabla$} \cdot\bA=0$ and \bA\ parallel to the direction of motion of the tube. The solution in spherical coordinates is then
\begin{equation}
\bA(\br,\br_Q,\bp_Q) = \bhvp {av_{Q\perp}\over r_Q r} \int_0^r dr' r'^2 \gamma_\perp \left({r'\over r_q}v_{Q\perp}\right) \,.
\label{eq:spherical}
\end{equation}
It is easy to verify that $\bB = (\bhr\times\bhvp){1\over r}
{\partial\over\partial r} (rA_\perp)$ is the same as in (\ref{eq:fvp}).

Similarly, the time component $A_0(\br,\br_Q,\bp_Q)$ is defined by
\begin{equation}
\bE = -\hbox{\boldmath$\nabla$}_{\br} A_0 = -a\gamma_\perp \left({r\over r_Q}v_{Q\perp}\right)\bhr\,,
\end{equation}
giving
\begin{equation}
A_0(\br,\br_Q,\bp_Q) = a\int_0^r dr' \gamma_\perp \left({r'\over r_Q} v_{Q\perp}\right)\,.
\label{eq:A^0}
\end{equation}
Both (\ref{eq:spherical}) and (\ref{eq:A^0}) can be explicitly integrated.

\section{Interpretation of the Laboratory Four Potential}

We first observe from (\ref{eq:spherical}) and (\ref{eq:A^0}) that for a pure flux tube the ``invariant" potential is
\begin{eqnarray}
U &=& A_0 - \bv \cdot \bA \,,\\
U &=& a \int_0^r dr' \gamma_\perp\left(v'_\perp\right) \left(1-v'^2_\perp\right) = a \int_0^r dr' \sqrt{1-v'^2_\perp}\,.
\end{eqnarray}
The resulting Lagrangian $L_{\rm string} = -U$ is exactly that of the string action\cite{ida,siminov} and the dynamics of a straight string with a massive spinless quark at one end (the other end being fixed) follows from the Lagrangian
\begin{equation}
L = -m\sqrt{1-v_Q^2} - U(\br=r_Q,\br_Q,p_Q)\,.
\label{eq:lagrange}
\end{equation}

The above Lagrangian reminds us that in a first quantized effective Hamiltonian all fields are to be evaluated at the quark and all derivatives refer to quark coordinates. The spinless quark case is straightforward since no derivations of fields appear and $U$ is well defined. The numerical methods of solving this problem are by now well established.\cite{collin,belgians} An example of the global spectroscopy is shown in Fig.~2. In this example the quarks at the ends of the flux tube are assumed massless.

\begin{figure}[t]
\centering
\epsfxsize=3in\hspace{0in}\epsffile{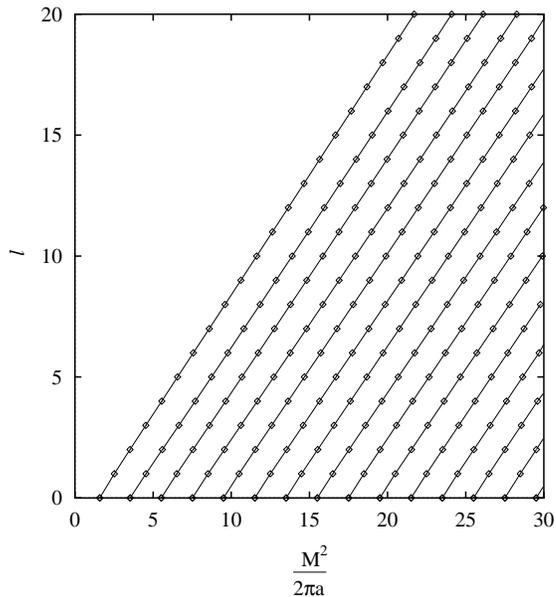}

\caption{The global Regge spectroscopy resulting from the numerical solution of (\protect\ref{eq:lagrange}) with massless quarks $(m=0)$ and two free ends.\protect\cite{collin,belgians} The vertical axis is orbital angular momentum and the meson mass $M$ is scaled so that the Regge slope is unity.}
\end{figure}

The potentials we have defined in (\ref{eq:spherical}) and (\ref{eq:A^0}) are unusual in that they depend on the velocity of the quark. This introduces an apparent uncertainty between field derivatives and quark derivatives. We therefore define the required quark derivative as
\begin{equation}
\lim_{\br\to\br_Q} \hbox{\boldmath$\nabla$}_{\br_Q} A_\mu(\br,\br_Q,\bp_Q) \equiv \lim_{\br\to\br_Q} \hbox{\boldmath$\nabla$}_{\br} A_\mu(\br,\br_Q,\bp_Q)\,.
\label{eq:q-deriv}
\end{equation}
This assumption ensures a smooth transition from field point to quark and in particular it guarantees that the Buchm\"uller electric tube will yield a pure Thomas spin-orbit energy when analyzed in the lab frame. The assumption (\ref{eq:q-deriv}) is satisfied if spatial deriviatives are taken at a constant instantaneous angular velocity.

\section{Effective Hamiltonian and Reduction}

The one-fermion Bethe-Salpeter equation can, without approximation, be expressed as the one-particle Salpeter equation\cite{long}
\begin{equation}
\Lambda_+ \left[ \hbox{\boldmath$\alpha$} \cdot (\bp-\bA) + \beta m+ A_0 - E \right] \Lambda_+  \psi = 0 \,,
\label{eq:salpeter}
\end{equation}
where the usual energy projection operators are defined by
\begin{equation}
\begin{array}{rcl}
\Lambda_\pm &=& E_0\pm H_0/ 2E_0 \,,\\
E_0 &=& \sqrt{p^2+m^2} \,,\\
H_0 &=& \hbox{\boldmath$\alpha$} \cdot \bp + \beta m\,.
\end{array}
\end{equation}
We suppress the	 subscript on $\br_Q$ and $\bp_Q$ since it is clear that such equations must involve quark coordinates. The effective Hamiltonian is then
\begin{equation}
H_{\rm eff} = H_0 + \Lambda_+ (A_0 - \hbox{\boldmath$\alpha$}\cdot\bA)\Lambda_+\,.
\label{eq:Heff1}
\end{equation}
We have assumed that the potentials have been promoted to Hermitian operations by suitable symmetrization. The use of the Salpeter equation avoids possible inconsistencies inherent in the Dirac equation due to the Klein paradox.\cite{obs} For the purpose of determining relativistic corrections either equation gives the same result. The proof of this statement follows from  the expectation identity
\begin{equation}
\left< \left[ E_0^{\pm1},\, F(r) \right] \right> \simeq 0 
\end{equation}
valid to leading order in $p^2/m^2$.
From this and the standard reduction of the Dirac equation we obtain the reduced (semi-relativistic) approximation,
\begin{equation}
\begin{array}{rcl}
H_{\rm eff} &\simeq&\displaystyle \left(m + {p^2\over 2m} - {p^4\over 8m^3}\right) + A_0 + {\nabla^2 A_0\over 8m^2} - {\bA\cdot\bp\over m}+ H_{\rm s.o.}\,,\\
H_{\rm s.o.} &=&\displaystyle {1\over 2m^2r} {dA_0\over dr} \bL\cdot\bS - {1\over m} (\hbox{\boldmath$\nabla$} \times \bA) \cdot \bS \,.
\end{array}
\label{eq:reduced}
\end{equation}

To explicitly evaluate this reduction we need to expand $A_\mu$ in powers of $v^2$ and evaluate the derivative at the quark coordinate using (\ref{eq:q-deriv}). For low quark velocities we find from the general expressions (\ref{eq:spherical}) and (\ref{eq:A^0}) that
\begin{eqnarray}
A_0(r,r_Q,p_Q) &\simeq& ar + {av_{Q\perp}^2\over 6r_Q^2} r^3 \,,\\
\bA(r,r_Q,p_Q) &\simeq& {a\over3} {v_{Q\perp}\over r_Q} r^2 \bhvp \,.
\end{eqnarray}
Using consistently the prescription (\ref{eq:q-deriv}) to evaluate the field derivatives we obtain
\begin{equation}
\begin{array}{rcl}
\nabla^2A_0 &\simeq& 2a/ r_Q\,,\\
\displaystyle {dA_0\over dr} &\simeq& a\,,\\
\hbox{\boldmath$\nabla$}\times\bA &\simeq& av_{Q\perp} (\bhr\times\bhvp)\,.
\end{array}
\end{equation}
Finally, using the fact that a heavy quark carries most of the orbital angular momentum
\begin{equation}
L \simeq mr_Q v_{Q\perp}\,,
\end{equation}
(\ref{eq:reduced}) becomes
\begin{equation}
H_{\rm eff} \simeq \left(m + {p^2\over m} - {p^4\over 8m^3}\right) + ar + {a\over 4m^2r} - {aL^2\over 6m^2r} - {a\over 2m^2r} \bL \cdot \bS\,.
\label{eq:Heff}
\end{equation}

\section{The Mechanical Relativistic Flux Tube Model}

The relativistic flux tube (RFT) model is based on the energy-momentum Lorentz transformation.\cite{collin,ted} In the co-rotating frame the electric flux tube is assumed to have constant energy per unit length equal to the tension $a$. In a frame where the tube rotates with angular velocity $\omega\equiv v_{Q\perp}/r_Q$ the energy of an infinitesimal element is then $adr/\sqrt{1-\omega^2r}$ and the total energy of the tube is
\begin{equation}
H_t(r) = a\int_0^r {dr'\over \sqrt{1-\omega^2 r'^2}} \,,
\end{equation}
which is the same as $A_0(r)$ in (\ref{eq:A^0}). Similarly, the angular momentum of the tube $\bL_t$ is
\begin{equation}
\bp_t \equiv {\bL_t\over\ r} = {a\omega\over r} \int_0^r {dr' r'^2 \over
\sqrt{1-\omega^2 r'^2}} (\bhr\times\bhvp) \,,
\end{equation}
which is the same as $\bA(r)$ from (\ref{eq:spherical}).

In the free Salpeter equation,\cite{long}
\begin{equation}
\Lambda_+ [ \hbox{\boldmath$\alpha$}\cdot\bp + \beta m -E\openone] \Lambda_+ \Psi = 0 \,,
\end{equation}
the confining interaction is introduced by the ``covariant tube substitution"\cite{ted}
\begin{equation}
p^\mu = (E, \, \bp) \to (E-H_t, \, \bp-\bp_t) \,,
\end{equation}
giving
\begin{equation}
\Lambda_+ [ \hbox{\boldmath$\alpha$} \cdot (\bp-\bp_t) + \beta m + H_t \openone - E \openone ] \Lambda_+ \psi = 0 \,.
\end{equation}
This is identical to (\ref{eq:salpeter}) since $H_t = A_0$ and $\bp_t = \bA$. We see that the assumption of pure chromo-electric field in the co-rotating frame is equivalent to the mechanical RFT model. 

\section{Conclusions}

Our main result is contained in the effective Hamiltonian (\ref{eq:salpeter}) and its reduced form (\ref{eq:Heff}). The starting point was Buchm\"uller's observation that a moving flux tube should be pure electric in the co-rotating frame. Although this makes very plausible the pure Thomas spin-orbit term and the lack of spin-spin interactions at large distance, it did not directly constitute a complete effective Hamiltonian. The present analysis, done in a non-rotating frame, extends Buchm\"uller's original conclusions to achieve a complete dynamical theory. In addition to the usual kinetic energy terms and the static linearly confining energy, there are three relativistic corrections which are dependent on the type of interaction. These are the last terms in (\ref{eq:Heff}): the Darwin, the $L^2$, and the Thomas terms. The Darwin term depends on the method of symmetrizing the Hamiltonian and thus is somewhat ambiguous. The $L^2$ and the Thomas spin-orbit terms are exactly what one expects from a more fundamental QCD approach.\cite{barch}

An additional significant observation is that the original effective Hamiltonian (\ref{eq:Heff1}) before any semi-relativistic approximations were made is exactly that of the ``mechanical relativistic flux tube model".\cite{collin,ted} This model was constructed by considerations of mechanical momentum and energy conservation of a flux tube with quarks at its ends. The relativistic treatment of the present case will also then have the desired Regge behavior (for massless quarks) of a Nambu string.\cite{ted}

Finally, we might mention that one can also start with the direct assumption that $\bB'=0$ on the quark.\cite{newpaper} This is another way to build in the Thomas spin-orbit interaction and by a suitable Lorentz transform method obtain a complete dynamical model. In this case the resulting model is relatively easy to solve because the Hamiltonian can be explicitly written in terms of canonical momenta and coordinates. Although it promises to be a useful model it does not contain direct reference to the flux tube properties. The relativistic corrections  turn out to be different than those in (\ref{eq:Heff}) and the Regge slope is the same as from scalar confinement.

\section*{Acknowledgments}

The contributions of my collaborators Ted Allen, Sinisa Veseli, and Ken\break Williams  are gratefully acknowledged. This work was supported in part by the U.S.~Department of Energy under Grant No.~DE-FG02-95ER40896 and in part by the University of Wisconsin Research Committee with funds granted by the Wisconsin Alumni Research Foundation.

\end{document}